\documentclass{cdclass}

\begin{document}

\title{Revisiting the paper "Simulating dynamical features of escape panic": What have we learnt since then?}
%\titlerunning{Short form of title} % if too long for running head

\author{
  Milad Haghani\authorlabel{1} \and 
  Enrico Ronchi\authorlabel{2} %\and
}
%short form of author list (without footnote symbol) for running head; may be using "et al." 
\authorrunning{M. Haghani \and E. Ronchi}
\institute{
  \authorlabel{1} University of New South Wales, Australia
  \authoremail{1}{milad.haghani@unsw.edu.au}
  \and
  \authorlabel{2} Division of Fire Safety Engineering, Lund University, Sweden
  \authoremail{2}{enrico.ronchi@brand.lth.se}
  %\and
}

\maketitle

\nolinenumbers
\begin{abstract}
The paper "Simulating dynamical features of escape panic" by Helbing, Farkas, and Vicsek, published over two decades ago in Nature, has left an indelible mark on the field of crowd dynamics. With nearly 3,000 citations to date, according to the Web of Science records, and significant influence, it has shaped the crowd dynamics field. This analysis investigates the overall influence of this paper through a variety of indicators, mapping its reach across research areas. The intellectual foundation of the paper is traced, examining the references cited. The terminological impact is also explored, showing how the paper made use of terms like "panic" and "herding". Moreover, the alignment of the assumptions of the paper with empirical evidence is discussed, finding discrepancies in key assertions about panic behaviour. The numerical simulations of the paper and observations have significantly influenced the field, such as for the case of the "faster-is-slower" phenomenon. The paper remains a key pillar in crowd dynamics, nevertheless, we advocate for a new course of the field shifting away from the terminology adopted in the paper and focusing more on empirical evidence. 
\end{abstract}

% e.g. self-driven particles; pedestrians; vehicles; animals;
% molecular motors; agents; traffic; crowds; swarms; experiments;
% models; microscopic; macroscopic; celular automata; software; open
% source

%%%%%%%%%%%%%%%%%%%%%%%%%%%%%%%%%%%%%%%%%%%%%%%%%%%%%%%%%%%%%%%%%%%
\section{Introduction}
\label{sec:intro}

More than two decades have passed since the publication of one of the most influential papers in the field of crowd dynamics authored by Helbing, Farkas, and Vicsek, which was featured in Nature (Helbing et al. \cite{helbing2000simulating}). Remarkably, this paper has obtained nearly 3,000 citations according to Web of Science and 6,000 citations according to Google Scholar up to now. Its influence extends beyond mere numerical counts, shaping the trajectory of crowd dynamics research in profound ways since its inception. The paper is not just the most cited paper in the field; it has also fundamentally shaped the discourse and development of crowd dynamics as a standalone discipline. This prompts us to consider how differently our understanding of crowd dynamics would have evolved had this groundbreaking paper not appeared on the pages of Nature. Has this influence predominantly steered the field in a positive direction, or have there been negative consequences? If the impact has been mixed, what aspects have been beneficial and what aspects have not? Our investigation operates on two distinct levels: a bibliometric analysis and a content-focused examination. We undertake several key inquiries: \\
\textbf{(1) Overall influence}: Through analysis of bibliometric indicators, we quantify the influence of the paper based on objective metrics, while also mapping the extent of influence that this paper has had and the areas of research to which its influence has permeated. \\
\textbf{(2) Intellectual Foundation}: We trace the intellectual lineage that led to the genesis of this paper by dissecting its references, and in some cases, the references within those references. By doing so, we aim to unveil the origins and foundations of concepts like \textit{escape panic} as referred by the authors. \\
\textbf{(3) Terminological Influence}: We explore how the language and terminology employed in this paper have shaped the broader lexicon of crowd dynamics. To gauge this influence, we analyse the language employed in the field prior to and subsequent to the publication of this paper. \\
\textbf{(4) Assumptions vs. Empirical Evidence}: Helbing et al. \cite{helbing2000simulating} outlined a set of assumptions characterising \textit{crowd (escape) panic}. Since then, numerous empirical studies have emerged, testing these assumptions through various experimental methods. We critically assess whether this accumulated body of evidence aligns with the conceptual characteristics laid out in the original paper. \\
\textbf{(5) Numerical Simulations and Experimental Replication}: Following the introduction of the social force modelling concept, Helbing et al. \cite{helbing2000simulating} conducted a series of numerical simulations that yielded noteworthy conclusions. These findings have permeated the crowd dynamics literature. We evaluate the extent to which subsequent experimental studies have investigated these numerical observations. \\
Overall, this study endeavours to provide an evidence-based revisiting of the influence, intellectual underpinnings, terminological evolution, and validation of assumptions and numerical findings within the referenced work. Such an assessment can help guide the trajectory of the field, offering insights into whether a conscious paradigm shift is necessary to address the issues associated with its influence. From here on, we refer to the paper as the HFV paper (after the initials of the authors’ surnames).

\section{Overall Influence}
\label{sec:influence}

\begin{figure}
  \centering
  \includegraphics[width=1\textwidth]{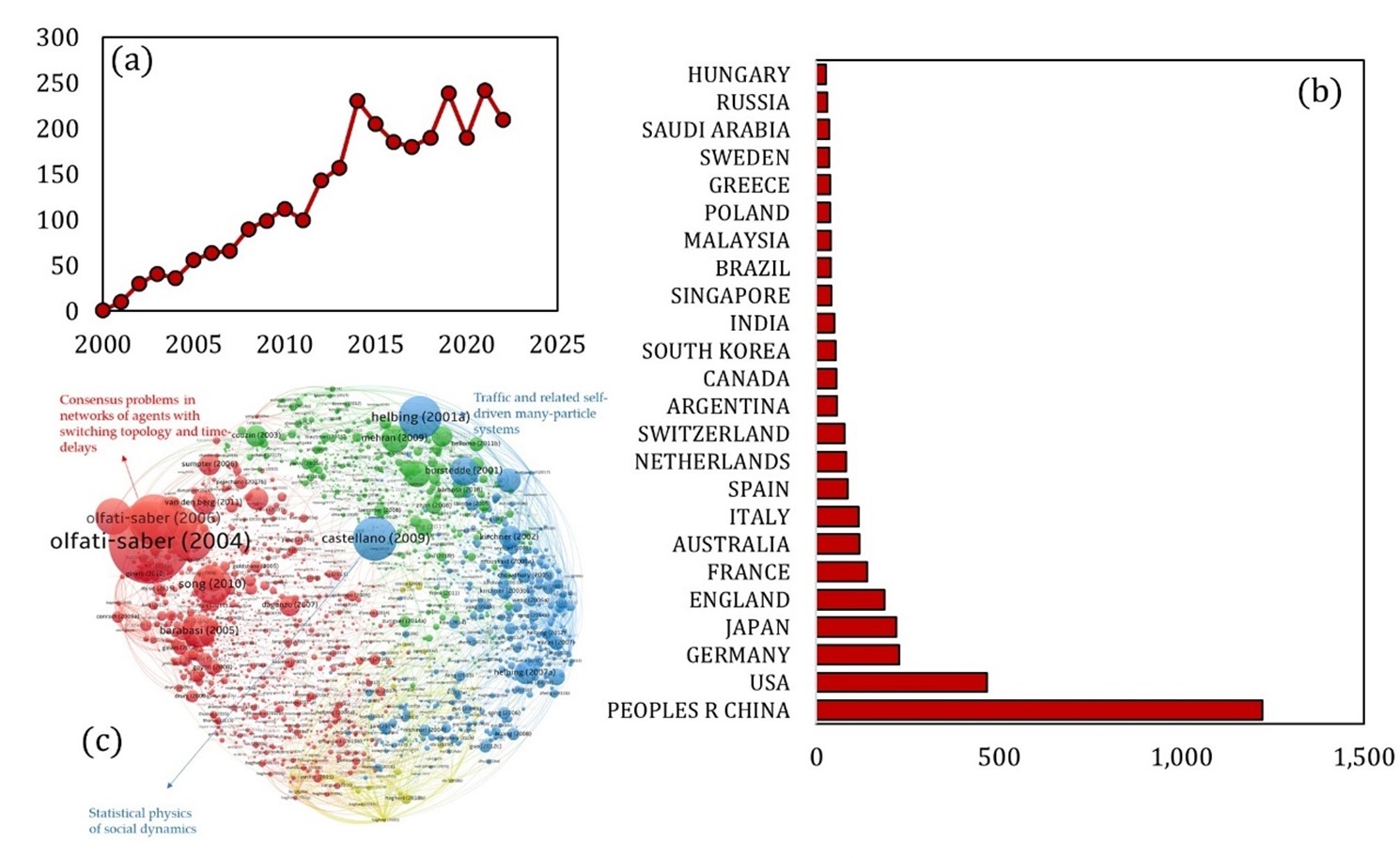} 
  \caption{The number of citing articles of HFV paper every year since its publication (part (a)), the countries of origin of the citing articles and the clusters of citing articles and their relative prominence (part (c)).}
  \label{fig:1} % unique label
\end{figure}

At the time of this analysis, we identified nearly 3,000 papers that have referenced the HFV paper, according to the Web of Science, making it the second most cited paper ever published in the crowd dynamics literature. It follows an earlier work by Helbing and Molnar \cite{helbing1995social} where they laid the foundation of the Social Force (SF) model of pedestrians and introduced the model for the first time. The rate of annual citations for the HFV paper exhibited a consistent linear increase since its publication and that trend continued until 2015, when there was a noticeable change in this growing trend (See \fref{fig:1}). This may coincide with a paradigm shift in crowd dynamics research around that year. Researchers began to shift their focus significantly towards experimental work, diverging from the predominant trend of numerical work up until that point.

One noteworthy aspect is the substantial portion of these citations originating from papers authored by researchers based in China, indicating the exceptional popularity and influence of this paper within this community. The majority of citing articles belong to the fields of Physics and Computer Science, with a notable concentration of these articles in the journal Physica A.
Within the field of crowd dynamics, the HFV paper has also left its mark on influential papers such as \cite{helbing2001traffic} and \cite{burstedde2001simulation} which introduced the concept of cellular automata modelling to the field in 2001.
To gain a deeper understanding of the areas where the HFV paper has exerted its influence, we conducted a document co-citation analysis among the reference lists of the nearly 3,000 citing papers. This analysis identifies clusters of references that are co-cited by these papers. In other words, it reveals groups of references that tend to appear together in the citations to the HFV paper by its citing articles. We identified ten such clusters, and by analysing the key phrases in the titles of these citing articles, we identified and categorised the themes of studies that have referenced the HFV paper. These themes are listed in \aref{app:1} in order of significance, showing how the influence of this paper has extended beyond the field of crowd dynamics. Additionally, \aref{app:1} provides a list of references that have been most frequently co-cited with the HFV paper.

\section{Intellectual Foundation}
\label{sec:foundation}

The HFV paper cites a total of twenty references. In our analysis, we do not intend to scrutinise every single reference, as not all appear to be equally fundamental in shaping its intellectual foundation. Notably, the initial references cited in the HFV paper have played a pivotal role in setting the stage and justifying the study and modelling of \textit{escape panic}. An intriguing observation concerning these initial references is that not all of them appear to support the argument; in fact, some seem to directly contradict the statements they accompany. For instance, the paper by Keating \cite{keating1982myth} is cited following a statement asserting \textit{"Sometimes this behaviour [panic] is triggered in life-threatening situations such as fires"}. The reference title \cite{keating1982myth}, \textit{The myth of panic}, and its content strongly suggests that it does not align with the statement, but rather points in the opposite direction. Similarly, the subsequent statement reads, \textit{"At other times, stampedes can arise during the rush for seats"}. One of the two references cited following this statement is the 1987 study by Johnson \cite{johnson1987panic}. However, the content and conclusions of Johnson’s paper appear to contradict the statement. This study \cite{johnson1987panic} analyses empirical evidence related to a tragic incident before a rock concert and reports \textit{“evidence showing that panic did not cause the death and injury of numerous young people"}. Furthermore, it highlights that post-disaster interviews and event reconstructions revealed no signs of stampede; instead, most competition stemmed from people's attempts to escape a crowd crush. This contradicts the notion that \textit{panic} can arise from people's rush for seats, as stated by HFV. Interestingly, in an editorial written by David Low about the HFV paper in the same issue of Nature, Johnson’s paper is inaccurately cited following this statement, \textit{"The consequences of crushing, trampling, and panic in crowds are well known"}. References \cite{keating1982myth} and \cite{johnson1987panic} are repeatedly cited by HFV throughout the paper without them offering support in favour of the arguments that they accompany.

Another example arises in the statement, \textit{"Panicking individuals tend to show maladaptive and relentless mass behaviour like jamming and life-threatening overcrowding, which has often been attributed to social contagion"}. The 1957 work of Quarantelli \cite{quarantelli1957behavior} is frequently cited following many of these statements. Published in 1957, it has been hard to trace this reference in contemporary databases. However, we managed to locate a more recent working paper by Quarantelli from 2001 \cite{quarantelli2001panic}, which we believe would naturally reflect the author's observations and reports from earlier stages of their career. The working paper is titled \textit{The Sociology of Panic} and it clearly indicates that the Quarantelli's findings could not have possibly been consistent with the portrayal presented in the HFV paper.
Taken together, these observations cast doubt on some of the foundational ideas upon which the HFV paper appears to rely. In fact, certain aspects of the work cited seem to fundamentally contradict key premises of the paper or at least reflect a possible inaccurate use of terminology.

\section{Terminological Influence}
\label{sec:terminology}

An examination of the text within the article reveals the prominent usage of two terms: \textit{panic} (mentioned 37 times) and \textit{herding} (mentioned 10 times). The notable prevalence of these terminologies within the field of pedestrian dynamics can be directly attributed to the influential impact of this article. As shown in \cite{feliciani2023crowd}, prior to the publication of this article, there was minimal mention of these two terms in the context of studies on crowds and evacuations.
Similarly, the terminology of \textit{crowd dynamics} (mentioned 3 times), which has been detected in the titles, abstracts, or keywords of more than 700 articles since 2000, was scarcely identifiable in any publications within the aforementioned domains before the publication of this article. It is reasonable to conclude that the nomenclature by which this field is recognised owes its existence to the terminological influence exerted by this article.

\section{Assumptions vs. Empirical Evidence}
\label{sec:assumptions}

The foundation of the HFV paper hinges on the concept of \textit{panic}, which it delineates through a set of defining characteristics. In this analysis, we prospectively reevaluate some of these key assumptions to determine whether they align with the empirical evidence that has emerged in the field since that publication. \textbf{\textit{People move or try to move considerably faster than normal}}.  This behaviour does not necessarily indicate panic per se. \textbf{\textit{Individuals start pushing, and interactions among people become physical in nature}}. Empirical evidence does not support the notion that people engage in competitive behaviour when responding to a life-threatening situation \cite{drury2009cooperation, bartolucci2021cooperative}. \textbf{\textit{At exits, arching and clogging are observed}}. Arching is not a particular characteristic of panic; rather, it is observable to some extent in laboratory experiments when a crowd of pedestrians is instructed to pass through a narrow bottleneck  \cite{uesten2023exploring}. \textbf{\textit{Jams build up}}. Similar to the previous point, the formation of a pedestrian traffic jam is linked to a situation where the inflow of pedestrians exceeds the capacity of a passage. \textbf{\textit{People show a tendency toward mass behaviour, that is, to do what other people do}}. This assumption is particularly important because it is readily operationalisable in evacuation models. In many cases, in fact, crowd models have been assessed on their ability to replicate such phenomenon. However, empirical evidence accumulated since the HFV paper suggests that the matter of imitation is complex. Social influence has been observed in several experimental studies, but there are several factors and variables affecting its extent, e.g., crowd size, familiarity with the environment, and perceived urgency \cite{haghani2019panic}. While there is evidence of some level of imitation regarding decision adaptation, it does not lead to blind herd-like behaviour where all individuals follow a single action (e.g., as visually depicted by the Figure 1 of the Editorial piece by David Low \cite{low2000following} on the HFV paper, marking one of the most influential illustrations produced in this field). In addition, empirical evidence has overwhelmingly suggested that humans take into consideration a range of factors in flights situations and not purely the decisions of others \cite{kinateder2018exit}. 

\section{Numerical Simulations and Experimental Replication}
\label{sec:replication}

The HFV article subjected its proposed crowd escape panic model to numerical testing, leading to recommendations that have significantly influenced both the field and public perception. Among these observations, the most consequential is the so-called \textit{faster-is-slower} phenomenon. The study considered the evacuation of a crowd of 200 agents through a 1m-wide exit while gradually increasing a parameter known as \textit{desired velocity}. The outcome revealed that as the desired velocity increased, system efficiency (in this context, the inverse of total evacuation time) initially improved. However, further increases in this parameter led to reduced system efficiency, implying longer evacuation times. This finding was translated into the recommendation that the most efficient way to evacuate and survive a crowded space with limited exit capacity, relative to occupancy, is for individuals to remain patient and keep lower speeds when passing through bottlenecks. Another suggestion was that asymmetrical placement of columns in front of exits can improve outflow. 
Several dimensions warrant consideration:
\textbf{\textit{(1) Empirical testing}}: Empirical studies have in some instances confirmed and in other contradicted the faster-is-slow and column phenomena \cite{haghani2019push, zuriguel2020contact, haghani2020empirical}. In a controlled experiment, participants were instructed to exit forcefully (without resorting to aggression), resulting in significantly higher flow efficiency compared to a calm and patient manner \cite{haghani2019push}. \textbf{\textit{(2) Model results need context}}: Numerical findings could be very parameter specific. According to Figure 1 (c) of the original paper, the minimum time required for 200 people to evacuate a room with a single 1m-wide exit is approximately 120s. Empirical observations indicate that this number of people may complete this evacuation in different (even faster) times depending on several factors. \textbf{\textit{(3) Desired velocity in non-free flow conditions}}: The concept of \textit{desired velocity} is hard to grasp physically, except in free-flow conditions. In scenarios with bottlenecks or obstacles, it remains unclear how different levels of desired flow correspond to various levels of escape motivation. It is possible that in reality, assuming that the model components can manifest physically in the real world (at least in terms of desired velocity), we are always within the monotonically decreasing limb of the graph in Figure 1(c) of the HFV article. This would imply that an increase in evacuation time does not manifest for typical (non-aggressive) forces applied by people at a bottleneck. This observation has had profound implications and has influenced both research studies that employed ants and mice as models for human crowds \cite{haghani2020empirical} as well as public perceptions regarding efficient evacuations. The main argument is that the panic concept assumed by the model may be misinterpreted as an encouragement to stay on the beginning of the left-hand side of the flow/density relationship, whereas we know that flow/density reaches a maximum with a speed which is at an optimal density value, and not at slow speed level.

\section{Conclusions}
\label{sec:conclusion}

Our analysis suggests that the HFV paper stands as one of the most fundamental and influential contributions to the field of crowd dynamics. It has not only profoundly shaped the terminologies employed within this field but has also significantly influenced the trajectory of research undertaken by scholars. It is evident that the introduction of the social force model of pedestrian dynamics marks a major paradigm shift in crowd research. This model introduced a paradigm that has spurred numerous developments in agent-based modelling of pedestrian traffic and beyond, establishing itself as a standard benchmark. Patterns of referencing to the article demonstrate its broad-reaching impact, reflecting its use as an intellectual foundation and source of inspiration in a multitude of domains. Our work highlights aspects that facilitate the interpretation of the HFV paper:
\textbf{(1) Model limitations}: There are a set of misconceptions regarding the model applicability. The model excels mostly when integrated with other layers of modelling, addressing various aspects of pedestrian behaviour and decision-making beyond the operational level and step-taking behaviour \cite{haghani2023crowd}.
\textbf{(2) Calibration procedure}: Presenting the model along with a detailed sensitivity analysis to identify critical model parameters would have helped understanding their impact on simulation output. A standardised calibration procedure could have facilitated future model applications. In the social force model paradigm, the forces lack physical manifestations, making them metaphorical or imaginary concepts. This renders them unmeasurable (though some have attempted to measure these forces in experimental settings \cite{zhao2019quantitative}), making it exceedingly difficult to link experimental observations to the model components and calibrate parameters using established procedures. This led to partial calibration attempts \cite{haghani2023crowd}.
\textbf{(3) Model validation}: The numerical case studies presented in the paper, including the \textit{faster-is-slower} phenomenon and the \textit{obstacle effect}, have been later scrutinised showing they may not hold for all scenarios. These findings have directed research efforts toward replicating them, in some instances relying on non-valid methods, such as animal experiments \cite{haghani2020empirical}. Relying on assumptions about panic, unsupported by references, has significantly influenced the direction of research in this field.

In summary, the HFV paper has played a pivotal role in the establishment of crowd dynamics as a research field. While we recognise the methodological and modelling benefits derived from this paper, we advocate for a paradigm shift in crowd dynamics research. It is essential to recognise that the mere publication of this paper in Nature should not serve as an unequivocal endorsement of the quest to delineate and explore the concept of \textit{crowd panic}. Thus, we call for a reimagining of the field, focused on empirical evidence to chart a new course for research in crowd dynamics.

\begin{contributions}
Milad Haghani: Conceptualization, Methodology, Software, Validation, Formal Analysis, Investigation, Writing – Original Draft, Visualization
Enrico Ronchi: Conceptualization, Methodology, Validation, Formal Analysis, Investigation, Writing – Review and Editing
\end{contributions}

\bibliographystyle{cdbibstyle} % mathematics and physical sciences
\bibliography{Haghani_Ronchi} % name your BibTeX data base

\begin{thebibliography}{10}
\providecommand{\url}[1]{{#1}}
\providecommand{\urlprefix}{URL }
\expandafter\ifx\csname urlstyle\endcsname\relax
  \providecommand{\doi}[1]{DOI~\discretionary{}{}{}#1}\else
  \providecommand{\doi}{DOI~\discretionary{}{}{}\begingroup \urlstyle{rm}\Url}\fi

\bibitem{helbing2000simulating}
Helbing, D., Farkas, I., Vicsek, T.: Simulating dynamical features of escape panic.
\newblock Nature \textbf{407}(6803), 487--490 (2000)

\bibitem{helbing1995social}
Helbing, D., Molnar, P.: Social force model for pedestrian dynamics.
\newblock Physical review E \textbf{51}(5), 4282 (1995)

\bibitem{helbing2001traffic}
Helbing, D.: Traffic and related self-driven many-particle systems.
\newblock Reviews of modern physics \textbf{73}(4), 1067 (2001)

\bibitem{burstedde2001simulation}
Burstedde, C., Klauck, K., Schadschneider, A., Zittartz, J.: Simulation of pedestrian dynamics using a two-dimensional cellular automaton.
\newblock Physica A: Statistical Mechanics and its Applications \textbf{295}(3-4), 507--525 (2001)

\bibitem{keating1982myth}
Keating, J.P.: The myth of panic.
\newblock Fire Journal \textbf{76}(3), 57--61 (1982)

\bibitem{johnson1987panic}
Johnson, N.R.: Panic at “the who concert stampede”: an empirical assessment.
\newblock Social Problems \textbf{34}(4), 362--373 (1987)

\bibitem{quarantelli1957behavior}
Quarantelli, E.L.: The behavior of panic participants.
\newblock Sociology and Social Research \textbf{41}, 187--194 (1957)

\bibitem{quarantelli2001panic}
Quarantelli, E.L.: Panic, sociology of  (2001)

\bibitem{feliciani2023crowd}
Feliciani, C., Corbetta, A., Haghani, M., Nishinari, K.: How crowd accidents are reported in the media: Lexical and sentiment analyses.
\newblock arXiv preprint arXiv:2309.14633  (2023)

\bibitem{drury2009cooperation}
Drury, J., Cocking, C., Reicher, S., Burton, A., Schofield, D., Hardwick, A., Graham, D., Langston, P.: Cooperation versus competition in a mass emergency evacuation: A new laboratory simulation and a new theoretical model.
\newblock Behavior research methods \textbf{41}(3), 957--970 (2009)

\bibitem{bartolucci2021cooperative}
Bartolucci, A., Casareale, C., Drury, J.: Cooperative and competitive behaviour among passengers during the costa concordia disaster.
\newblock Safety science \textbf{134}, 105055 (2021)

\bibitem{uesten2023exploring}
Uesten, E., Schumann, J., Sieben, A.: Exploring the dynamic relationship between pushing behavior and crowd dynamics.
\newblock Collective Dynamics \textbf{8}, 1--29 (2023)

\bibitem{haghani2019panic}
Haghani, M., Cristiani, E., Bode, N.W., Boltes, M., Corbetta, A.: Panic, irrationality, and herding: three ambiguous terms in crowd dynamics research.
\newblock Journal of advanced transportation \textbf{2019} (2019)

\bibitem{low2000following}
Low, D.J.: Following the crowd.
\newblock Nature \textbf{407}(6803), 465--466 (2000)

\bibitem{kinateder2018exit}
Kinateder, M., Comunale, B., Warren, W.H.: Exit choice in an emergency evacuation scenario is influenced by exit familiarity and neighbor behavior.
\newblock Safety science \textbf{106}, 170--175 (2018)

\bibitem{haghani2019push}
Haghani, M., Sarvi, M., Shahhoseini, Z.: When ‘push’does not come to ‘shove’: Revisiting ‘faster is slower’in collective egress of human crowds.
\newblock Transportation research part A: policy and practice \textbf{122}, 51--69 (2019)

\bibitem{zuriguel2020contact}
Zuriguel, I., Echeverr{\'\i}a, I., Maza, D., Hidalgo, R.C., Mart{\'\i}n-G{\'o}mez, C., Garcimart{\'\i}n, A.: Contact forces and dynamics of pedestrians evacuating a room: the column effect.
\newblock Safety science \textbf{121}, 394--402 (2020)

\bibitem{haghani2020empirical}
Haghani, M.: Empirical methods in pedestrian, crowd and evacuation dynamics: Part i. experimental methods and emerging topics.
\newblock Safety science \textbf{129}, 104743 (2020)

\bibitem{haghani2023crowd}
Haghani, M., Sarvi, M.: Crowd model calibration at strategic, tactical, and operational levels: Full-spectrum sensitivity analyses show bottleneck parameters are most critical, followed by exit-choice-changing parameters.
\newblock Transportation Letters pp. 1--28 (2023)

\bibitem{zhao2019quantitative}
Zhao, Y., Lu, T., Su, W., Wu, P., Fu, L., Li, M.: Quantitative measurement of social repulsive force in pedestrian movements based on physiological responses.
\newblock Transportation research part B: methodological \textbf{130}, 1--20 (2019)

\end{thebibliography}

\appendix
\section{Appendix 1}
\label{app:1}

Clusters of topics and research areas where the HFV paper has been cited, based on a document co-citation analysis of the citing articles of the HFV paper: \\
\textbf{Cluster 1}: pedestrian evacuation; social force model; bidirectional pedestrian flow; new lattice; visual field. \\
\textbf{Cluster 2}: crowd space; predictive crowd analysis technique; dense crowd; emotional contagion; personality trait. \\
\textbf{Cluster 3}: jamming transition; pedestrian evacuation; fire evacuation model; pedestrian dynamics; self-driven particle. \\
\textbf{Cluster 4}: collective escape; pedestrian counter flow; emergency escape; pedestrian contact force; pedestrian dynamics. \\
\textbf{Cluster 5}: collective motion; self-propelled particle; collective decision; nonlinear dynamics. \\
\textbf{Cluster 6}: pedestrian trajectory prediction; evacuation crowd dynamics; human behaviour; social interaction. \\
\textbf{Cluster 7}: dynamic decision behaviour; information service; discrete opinion dynamics; critical market crash. \\
\textbf{Cluster 8}: modelling pedestrian crowd; panic condition; crowd guidance; disaster evacuation; aircraft disembarking. \\
\textbf{Cluster 9}: switching phenomena; financial market; correlated randomness; trend switching processes; trend switching. \\
\textbf{Cluster 10}: granular material; clogging transition; horizontal hopper; hopper angle; large reflective obstacle. \\

\textbf{References most frequently co-cited by the HFV paper}: \\
1104 Co-citations: Helbing D, 1995, PHYS REV E, V51, P4282, \\ DOI 10.1103/PhysRevE.51.4282 \\
565 Co-citations: Burstedde C, 2001, PHYSICA A, V295, P507, \\ DOI 10.1016/S0378-4371(01)00141-8 \\
392 Co-citations: Kirchner A, 2002, PHYSICA A, V312, P260, \\ DOI 10.1016/S0378-4371(02)00857-9 \\
380 Co-citations: Helbing D, 2005, TRANSPORT SCI, V39, P1, \\ DOI 10.1287/trsc.1040.0108 \\
306 Co-citations: Helbing D, 2007, PHYS REV E, V75, P0, \\ DOI 10.1103/PhysRevE.75.046109 \\
291 Co-citations: Helbing D, 2001, REV MOD PHYS, V73, P1067, \\ DOI 10.1103/RevModPhys.73.1067 \\
285 Co-citations: Hughes RL, 2002, TRANSPORT RES B-METH, V36, P507, \\ DOI 10.1016/S0191-2615(01)00015-7 \\
270 Co-citations: Reynolds C.W., 1987, ACM SIGGRAPH COMPUTE, V21, P25, \\ DOI 10.1145/37402.37406 \\
268 Co-citations: Moussaid M, 2011, P NATL ACAD SCI USA, V108, P6884, \\ DOI 10.1073/pnas.1016507108 \\
261 Co-citations: Helbing D, 2002, PEDESTRIAN AND EVACUATION DYNAMICS, V0, P21 \\
229 Co-citations: Muramatsu M, 1999, PHYSICA A, V267, P487, \\ DOI 10.1016/S0378-4371(99)00018-7 \\
207 Co-citations: Vicsek T, 1995, PHYS REV LETT, V75, P1226, \\ DOI 10.1103/PhysRevLett.75.1226 \\
201 Co-citations: Helbing D, 2000, PHYS REV LETT, V84, P1240, \\ DOI 10.1103/PhysRevLett.84.1240 \\
193 Co-citations: Kirchner A, 2003, PHYS REV E, V67, P0, \\ DOI 10.1103/PhysRevE.67.056122 \\
181 Co-citations: Henderson LF, 1971, NATURE, V229, P381, \\ DOI 10.1038/229381a0 \\
169 Co-citations: Treuille A, 2006, ACM T GRAPHIC, V25, P1160,  \\DOI 10.1145/1141911.1142008 \\
167 Co-citations: Helbing D, 2003, PHYS REV E, V67, P0,  \\DOI 10.1103/PhysRevE.67.067101 \\
166 Co-citations: Hughes RL, 2003, ANNU REV FLUID MECH, V35, P169, \\ DOI 10.1146/annurev.fluid.35.101101.161136 \\
163 Co-citations: Zheng XP, 2009, BUILD ENVIRON, V44, P437, \\ DOI 10.1016/j.buildenv.2008.04.002 \end{document}